\documentclass[fleqn,usenatbib]{mnras}

\usepackage{newtxtext,newtxmath}

\usepackage[T1]{fontenc}

\DeclareRobustCommand{\VAN}[3]{#2}
\let\VANthebibliography\thebibliography
\def\thebibliography{\DeclareRobustCommand{\VAN}[3]{##3}\VANthebibliography}

\usepackage{graphicx}	
\usepackage{amsmath}	
\usepackage{float}
\usepackage{graphicx}
\usepackage{epsfig}
\usepackage{lineno}
\usepackage{url}
\usepackage{cleveref}
\usepackage[all]{hypcap}

\title[Challenges to Remote Sensing]{Fundamental Challenges to Remote Sensing of Exo-Earths}

\author[Paradise et al.]{
Adiv Paradise,$^{1}$\thanks{E-mail: paradise@astro.utoronto.ca}
Kristen Menou,$^{1,2,3}$
Christopher Lee$^{3}$
and Bo Lin Fan$^{1}$
\\
$^{1}$David A. Dunlap Department of Astronomy \& Astrophysics, University of Toronto, Toronto, ON M5S 3H4, Canada\\
$^{2}$Physics \& Astrophysics Group, Department of Physical and Environmental Sciences, University of Toronto, Scarborough, ON M1C 1A4, Canada\\
$^{3}$Department of Physics, University of Toronto, Toronto, ON M5S 1A7
}

\date{Accepted XXX. Received YYY; in original form ZZZ}

\pubyear{2021}

\begin{document}
\label{firstpage}
\pagerange{\pageref{firstpage}--\pageref{lastpage}}
\maketitle

\begin{abstract}
Inferring the climate and surface conditions of terrestrial exoplanets in the habitable zone is a major goal for the field of exoplanet science. This pursuit will require both statistical analyses of the population of habitable planets as well as in-depth analyses of the climates of individual planets. Given the close relationship between habitability and surface liquid water, it is important to ask whether the fraction of a planet's surface where water can be a liquid, $\chi_\text{hab}$, can be inferred from observations. We have produced a diverse bank of 1,874 3D climate models and computed the full-phase reflectance and emission spectrum for each model to investigate whether surface climate inference is feasible with high-quality direct imaging or secondary eclipse spectroscopy. These models represent the outcome of approximately 200,000 total simulated years of climate and over 50,000 CPU-hours, and the roughly-100 GB model bank and its associated spectra are being made publicly-available for community use. We find that there are correlations between spectra and $\chi_\text{hab}$ that will permit statistical approaches. However, spectral degeneracies in the climate observables produced by our model bank indicate that inference of individual climates is likely to be model-dependent, and inference will likely be impossible without exhaustive explorations of the climate parameter space. The diversity of potential climates on habitable planets therefore poses fundamental challenges to remote sensing efforts targeting exo-Earths.
\end{abstract}




\section{Introduction}
Identifying and characterizing both habitable planets and their uninhabitable counterparts will require not only detailed study of individual climates \citep{Arney2018,Ramirez2018,Quanz2019}, but broad statistical analyses of the exoplanet population itself \citep{Bean2017,Checlair2019}. We take a systematic approach to this problem, motivated by the realization in recent years that the habitable zone is host to a very wide range of climates. A planet's location in its parent star's assumed `habitable zone' is an insufficient guarantee of habitability, or even a temperate climate \citep{Ramirez2018a}. Planets may undergo Venus-like runaway warming \citep{Kasting1988,Nakajima1992}, freeze over entirely into snowball states \citep{Budyko1969,Caldeira1992}, or lose most of their atmospheres \citep{Lammer2008}. Among planets with more-temperate climates, they might be relatively dry \citep{Abe2011}, entirely water-covered \citep{Kite2018}, mostly frozen \citep{Paradise2017}, be tidally-locked \citep{Pierrehumbert2011}, or their atmosphere may have a very different mean molecular weight \citep{Koll2019}. Other factors such as obliquity \citep{Williams1997}, eccentricity \citep{Bolmont2016}, atmospheric mass \citep{Vladilo2013,Paradise2021}, the volcanic outgassing rate \citep{walker81}, plate tectonics \citep{Foley2019}, and rotation rate \citep{Abbot2018} can all affect the climates found in the habitable zone. Any inference technique that aims to be able to robustly separate temperate, habitable Earth-like climates from other climates must therefore contend with the possibility that all possible climates have not yet been explored, and a climate may exist which confounds that technique. Without any prior constraints on which climates may be most prevalent, this poses a serious challenge to the field.


Several techniques for identifying Earth-like climates have been proposed. The spectral shape of the broadband visible-light albedo has been proposed as a way to distinguish snowball and temperate climates \citep{Cowan2011}, and it has been proposed that multiple photometric bands could be used together to distinguish temperate climates in color-color space \citep{Krissansen-Totton2016}. If time-resolved observations are available, broadband photometric and polarimetric phase curves may be sensitive to the presence of liquid water oceans and land \citep{Cowan2013,Fujii2017,Trees2019}. Other efforts have been proposed using high-resolution spectra and phase curves to probe atmospheric composition \citep{Wolf2019, Chen2019}, colors to identify qualitative planet type \citep{Madden2018}, and albedo measurements to probe the existence of atmospheres \citep{Mansfield2019,Koll2019}. Yet another possibility that has been explored extensively is the role that polarization and ocean glint could play in helping to distinguish and identify surface types \citep[e.g.][]{McCullough2006, Stam2008, Williams2008, Robinson2010, Zugger2010, Emde2017}, though these techniques also face challenges from degeneracies and false positives \citep{Zugger2010,Zugger2011,Cowan2012}. It is possible that time-resolved or phase-resolved observations may help to break those degeneracies \citep{Visser2015,Lustig-Yaeger2018}. However, all of these analyses either used lower-dimensional models that cannot capture the spatial heterogeneity of realistic climates, or relied on a relatively small number of model climates, making it difficult to quantify their robustness to model parameterizations and assumptions. This motivates us to apply full-phase reflection and emission spectroscopy, as might be measured through differential secondary eclipse spectroscopy \citep{Richardson2007,Swain2008}, to a very large ensemble of 3D climate models spanning a large range of Earth-like planets, as a first step towards exploring and quantifying the impact that diversity in the population of Earth-like exoplanets might have on climate retrievals.

In this study, we used PlaSim, a 3D Earth climate model of intermediate complexity, to systematically model 1,874 climates of Earth-like planets in the habitable zones of Sun-like stars, varying surface pressure, rotation rate, CO$_2$ partial pressure, land fraction and distribution, and instellation. PlaSim is a coupled model that includes the atmosphere, ocean, and land \citep{Fraedrich2005}, and has been used before to study a range of climates \citep[e.g.][]{Lucarini2013,Paradise2017,Checlair2017}. We then use the radiative transfer code SBDART \citep{disort,Ricchiazzi1998,modtran} to produce disk-averaged reflected light and emission spectra of each model at full phase at 20 cm$^{-1}$ resolution, from 350 nm to 80 $\mu$m. We use reflection and emission spectra because transmission spectroscopy is primarily sensitive to the upper atmosphere \citep{Kreidberg2018,hightransmission1,hightransmission2}, whereas spectra obtained through direct imaging or secondary eclipse spectroscopy may more-directly probe surface conditions and climate \citep{Richardson2007,Swain2008}. The climates represented by our model bank range from cold snowball planets to warm climates with mean surface temperatures over 50 $^\circ$C. We have also included in our analysis cold, dry planets with various albedos, as a dry planet with a very reflective surface might superficially resemble a snowball planet, and such a planet with a dark surface might resemble a temperate ocean-covered planet \citep{Rushby2020}. To permit comparison between ocean-covered plants and drier land planets, we focus our investigation on the fraction of the planet's surface with temperatures that could support liquid water, or fractional habitability $\chi_\text{hab}$ \citep{Spiegel2008}, regardless of whether liquid water is present there in significant quantities in our models. We then use these synthetic observables to investigate the degree to which fractional habitability can be inferred from reflected-light spectra, both by revisiting previously-proposed correlations and examining potential new correlations.

\section{Methods}\label{obsec:methods}
\subsection{Modeling Pipeline}
\subsubsection{Climate Model}
We use PlaSim, a global climate model of intermediate complexity \citep{Fraedrich2005}, to simulate a wide range of planetary climates. PlaSim uses a spectral dynamical core to solve the primitive equations for vorticity, temperature, pressure, divergence, and specific humidity in a three-dimensional model atmosphere. PlaSim includes moist processes such as shallow and deep convection, diagnostic cloud formation, evaporation, precipitation, and re-evaporation of falling precipitation. Radiation is implemented through a three-band model, with two shortwave bands and one longwave band. PlaSim includes Rayleigh scattering and absorption by CO$_2$, H$_2$O, and O$_3$. Ozone absorption is prescribed, while CO$_2$ and H$_2$O absorption depend on their concentrations. PlaSim's atmosphere is coupled to both land and sea surfaces. The land surface includes a bucket hydrology, where each grid cell can fill with water up to a certain capacity. Excess surface water is treated as runoff, and advected to the continental margin according to average topography, creating a river system. Surface water and runoff both contribute to evaporation. Several land types can be prescribed, and their thermodynamic and reflective properties are modified by water saturation and the presence of snow or ice. The ocean is modeled as a 50-meter-thick mixed-layer slab, with sea ice computed thermodynamically. We use PlaSim with 10 vertical layers in its T21 configuration, with 32 latitudes and 64 longitudes, resulting in a resolution of roughly $5.6^\circ\times5.6^\circ$. For each model, we run for a minimum of 75 years, and then check for energy balance equilibrium. We compute the decadal-average energy balance at the surface and top of the atmosphere for each year, and then compute the 5-year mean rate of change in that energy balance for each year. We define the model as having reached energy balance when the 30-year average of the 5-year mean energy balance drift is less than $4\times10^{-4}$ W/m$^2$, which is less than $5.0\times10^{-7}$ times the lowest instellation considered in our model bank. We use this criterion because inter-annual variability can mean that a strict instantaneous energy balance requirement is never satisfied even when the overall climate state has been at equilibrium for several decades Most models reach this point after 100--300 years, depending on model parameters and initial conditions.

We wanted to take into consideration the possibility of confusion by thick Rayleigh-scattering atmospheres, but PlaSim's Rayleigh scattering is parameterized to a surface pressure of 1 atm. We therefore modified PlaSim's radiation model to introduce a dependence on atmospheric mass. We describe these modifications in more detail in another study \citep{Paradise2021}, but give a summary here. Rather than optical depth $\tau$ or transmittance $T$, which can be reasonably easily scaled with atmospheric mass, PlaSim's radiation model computes the reflectance $R$ of an atmospheric layer due to scattering. Reflectance is related to transmittance and optical depth, where $R=1-T$ and $T=e^{-\tau}$.The scattering optical depth is directly proportional to the mass of scatterers in the column, so we scale PlaSim's prescribed reflectance due to Rayleigh scattering $R_0$ by the atmospheric column mass $p_s/g$:
\begin{equation}
R = 1-\exp\left(\frac{p_s}{p_{s,\oplus}}\frac{g_\oplus}{g}\ln({1-R_0})\right)
\end{equation}
This modified scheme produces the same amount of scattering in the limit of Earth's atmosphere, and produces top-of-atmosphere fluxes that scale similarly to those computed with SBDART for atmospheric columns of varying masses, as described in \citet{Paradise2021}.

We also slightly modified PlaSim's vertical discretization. PlaSim's native configuration has vertical layers spaced mostly-linearly in pressure, which means that increasing surface pressure results a higher mid-level pressure for the top layer of the atmosphere. Since some radiative processes depend more on the optical depth from the top of the atmosphere than from the ground, we pin the top layer's mid-level pressure at approximately 50 hPa, and scale the rest of the atmosphere to retain the same functional form, but stretched from the ground to the new model top. In the limit of a 1-bar atmosphere, this vertical discretization is identical to PlaSim's default scheme, in which model levels are at constant values of $\sigma$, where $\sigma=p/p_s$.  

\subsubsection{Comparing planets with $\chi_\text{hab}$}

\begin{table*}
\centering
\begin{tabular}{p{1cm} p{2.5cm} p{4cm} p{4cm} p{3.5cm}}
\hline\hline \\
\# & Surface Type & Model parameter & pCO$_2$ & Insolation \\
\hline\hline \\
179 & Earth & pN$_2$ = 0.1--10 atm & pCO$_2$=360 $\mu$bar & $S$ = 1100--1550 W/m$^2$ \\ \hline
10 & Aquaplanet & pN$_2$ = 1--10 atm & pCO$_2$=360 $\mu$bar & $S$ = 1300 W/m$^2$ \\ \hline
964 & Aquaplanet & 0.1--10 day rotation & pCO$_2$=$10^{-3}$--$10^2$ mbar & $S$ = 1225--1275 W/m$^2$\\ \hline
395 & $f_\text{land}$ = 0.1--0.8 & 5 random distributions per $f_\text{land}$ & pCO$_2$=$10^{-3}$--$10^2$ mbar & $S$ = 1250 W/m$^2$ \\ \hline
20 & Aquaplanet & --- & pCO$_2$=$10^{-3}$--$10$ mbar & $S$ = 1350 W/m$^2$ \\ \hline
100 & Earth & Cold Start, pN$_2$ = 1 bar  & pCO$_2$=$10^{-3}$--$10$ mbar & $S$ = 800--1400 W/m$^2$ \\ \hline
100 & Aquaplanet & Cold Start, pN$_2$ = 1 bar & pCO$_2$=$10^{-3}$--$10$ mbar & $S$ = 800--1400 W/m$^2$ \\ \hline
99 & Aquaplanet & Cold Start, pN$_2$ = 8 bar & pCO$_2$=$10^{-3}$--$10$ mbar & $S$ = 800--1400 W/m$^2$ \\ \hline
8 & Desert planet & Albedo=0.1--0.8 & pCO$_2$=500 $\mu$bar & $S$ chosen to produce 220 K effective temperature \\ \hline\hline
\end{tabular}
\caption{Summary of the parameter sweeps performed in our ensemble of 1,874 models. For experiments where multiple variables were varied, they were varied together to form a grid of models. Obliquity was zero for all but the models with Earth-like continents, which had Earth-like obliquity as well. We do not include obliquity in any of our synthetic spectra.}\label{obtable:models} 
\end{table*}

The range of climates in our model bank is diverse enough that measures such as mean surface temperature, sea ice fraction, and maximum surface temperature are inadequate points of comparison. Two qualitatively different climates could have identical mean surface temperatures, yet one might be more habitable than the other. A planet with a moderate land fraction, most of which is near the equator, might have a larger sea ice fraction than an aquaplanet that is equivalently habitable, as a larger fraction of the former's oceans are ice-covered. Fractional habitability, $\chi_\text{hab}$, however is an alternative quantity that can be used to compare disparate climates. A planet's fractional habitability is simply the fraction of the planet's surface whose surface temperature permits liquid water \citep{Spiegel2008}. As none of our models exceed the water boiling point, we simplify the definition to the fraction of the surface that is above 273.15 K. We therefore define a binary mask, $H(\theta,\phi)$, where $\theta$ is latitude and $\phi$ is longitude, such that
\begin{equation}
H(\theta,\phi) =
\begin{cases}
1 & \text{if }\,\,T_s(\theta,\phi) \geq 273.15\,\,\text{K} \\
0 & \text{if }\,\,T_s(\theta,\phi) < 273.15\,\,\text{K}
\end{cases}
\end{equation}
The fractional habitability, $\chi_\text{hab}$, is thus the spatial average of $H(\theta,\phi)$:
\begin{equation}
    \chi_\text{hab} = \frac{\sum\limits_{j=1}^{N}\sum\limits_{k=1}^{M}H(\theta_k,\phi_j)A(\theta_k,\phi_j)}{\sum\limits_{j=1}^{N}\sum\limits_{k=1}^{M}A(\theta_k,\phi_j)}
\end{equation}
where $N$ is the number of longitudes, $M$ is the number of latitudes, and $A(\theta_k,\phi_j)$ is the surface area of grid cell ($j$,$k$). When we compare $\chi_\text{hab}$ between models, we use the annual average of $\chi_\text{hab}$. 

\subsubsection{SBDART}

\begin{figure*}
\begin{center}
\includegraphics[width=6in]{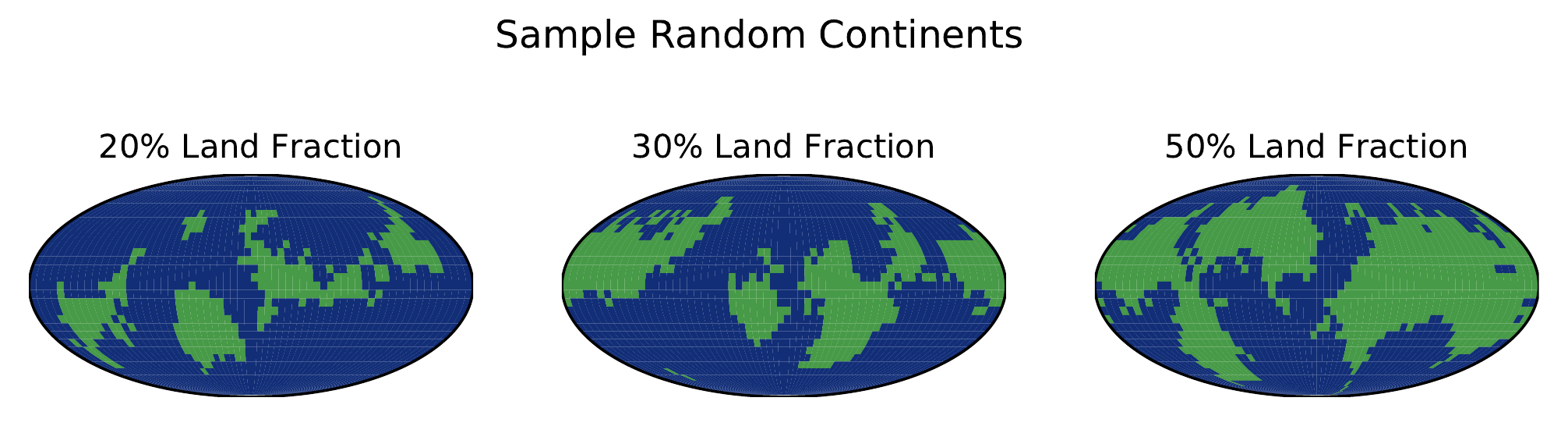}
\end{center}
\caption{Mollweide projections of three examples of land distributions produced by our random continent-generation algorithm, for land fractions of 0.2, 0.3, and 0.5. 5 such distributions were produced for each land fraction, for 40 total land configurations in this part of our experiment.}\label{obfig:continents}
\end{figure*}

We use the column radiative transfer model SBDART (Santa Barbara DISORT Atmospheric Radiative Transfer \citep{disort,Ricchiazzi1998,modtran}) to compute the reflectance and emission spectrum of each model. SBDART computes plane-parallel radiative transfer in both clear and cloudy atmospheres, and includes spectral models for a variety of surface types, including ocean, soil, and snow \citep{disort,Ricchiazzi1998,modtran}. We use the MODTRAN-3 input solar spectrum, which gives 20 cm$^{-1}$ resolution, and specify the atmospheric column properties directly from PlaSim model output. We compute radiative fluxes from 350 nm to 80 $\mu$m. We use the solar zenith angle as the observer viewing angle for the equatorial viewing geometry (giving maximum disk visibility), and for the polar viewing geometry place the observer above the North pole. Because SBDART assumes a horizontally-homogeneous plane-parallel atmosphere, we compute radiative transfer with SBDART independently for every PlaSim grid cell, specifying the solar zenith angle and the observer's viewing angle. 

For land surfaces, we use a weighted average of SBDART's sand and water models, with 5\% water. For our cold, dry planets with prescribed reflectivities, we specify a gray surface with uniform albedo. SBDART's default snow reflectance model is not defined in the portion of the near-infrared where reflected light is still stronger than emitted light, so instead we prescribe our own snow reflectance model, computed as a weighted average of snow reflectance models from the JPL ECOSTRESS spectral library \citep{aster,ecostress}. We use 30\% each of coarse-grained, medium-grained, and fine-grained snow in equal amounts, and then include 10\% pure ice. The resulting reflectance spectrum is included in \autoref{obfig:iceblends}. We also considered four other snow reflectance models: sea ice with melt pools (a 20\% contribution from liquid water reflectivity), frost, frost with 10\% andesite, and coarse-grained snow with 10\% andesite. Andesite is a grayish igneous rock whose composition roughly matches that of the continental crust on Earth \citep{andesite,aster,ecostress}. We use our snow spectral model for all grid cells whose surfaces are either entirely sea ice or have snow and an albedo in PlaSim greater than 0.2, the default ground albedo in PlaSim. Presence of snow and an albedo above 0.2 indicates widespread snow coverage---PlaSim modifies surface albedo according to the average snow depth in the cell, such that a very small amount of snow does not increase the albedo significantly. For oceanic grid cells whose surface is only partly covered in sea ice, we use a specification that is a combination of the spectral properties of snow and seawater.

We use PlaSim's output data to specify the column temperature, pressure, water vapor distribution, and clouds. SBDART permits as input both the liquid water path and frozen water path, but in the absence of the latter uses the CCM3 cirrus model to compute effective particle radii and scattering for clouds at low temperatures \citep{Ricchiazzi1998,ccm3-cirrus}. Because PlaSim does not report the liquid water path stored in clouds and precipitates excess humidity (neglecting cloud storage), we assume that the cloud liquid water path in a cloudy layer is simply the layer's total liquid water path. Our results therefore represent an upper limit on the optical depth from clouds in PlaSim. We also assume SBDART's default for water droplet effective radius (8 $\mu$m), as this is also not computed by PlaSim. PlaSim's output is normally time-averaged, and because clouds are transient to most layers, that would tend to reduce the cloud fraction in a given model cell. We therefore slightly modify PlaSim to also output snapshots with no time-averaging, and use those snapshots as input for SBDART instead of time-averaged data. 

\subsubsection{Integration and Postprocessing}

Once we have computed the reflection and emission spectrum for every grid cell for the desired viewing angles, we combine the spectra from the observer-facing hemisphere into a single spectrum. This combined spectrum, $F_\lambda(\Phi_{\text{obs}})$, is simply a weighted average of the component spectra, $F_\lambda(\theta,\phi,\Phi_{\text{obs}})$ using the area of the cell $A(\theta,\phi)$, where $\theta$ and $\phi$ are latitude and longitude and $\Phi_{\text{obs}}$ is the projection angle away from the normal direction:
\begin{equation}
F_\lambda(\Phi_{\text{obs}}) = \frac{\sum\limits_{\theta,\phi} F_\lambda(\theta,\phi,\Phi_{\text{obs}})A(\theta,\phi)\cos\Phi_{\text{obs}}}{\sum\limits_{\theta,\phi}A(\theta,\phi)\cos\Phi_{\text{obs}}}
\end{equation}

\subsection{Modeling Strategy}

\begin{figure*}
\begin{center}
\includegraphics[width=6in]{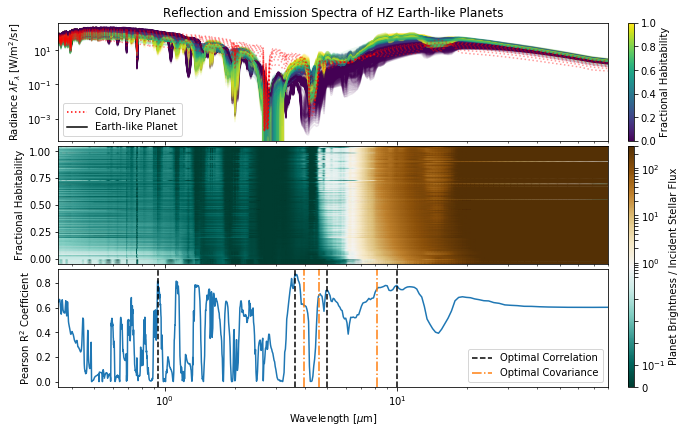}
\end{center}
\caption{High-resolution spectra for each of our 1,874 models. The top panel shows the radiance of each model from 350 nm to 80 $\mu$m. The middle panel shows the color-coded ratio of the planet's radiance to the incident stellar flux, as a function of wavelength and fractional habitability, $\chi_\text{hab}$, such that the `most-habitable' planets are at the top of the plot, and the `least-habitable' at the bottom. Wavelengths at which $\chi_\text{hab}$ varies with normalized radiance the most are particularly interesting for observational characterization. This is highlighted in the bottom panel, where the square of the Pearson's $R$ correlation coefficient between $\chi_\text{hab}$ and the log normalized radiance is shown for each wavelength. The midpoint of the middle panel's colorbar indicates the point at which thermal emission outweighs reflected light.  Models with $\chi_\text{hab}$=0 or $\chi_\text{hab}$=1 are included in the middle panel as small extensions above and below the 0--1 range, sorted by mean annual surface temperature. We have indicated the wavelengths of the 4 strongest local maxima in the correlation coefficient with vertical dashed lines in the bottom panel, and local maxima of the covariance are shown with orange dash-dotted lines. Note that despite obvious trends, there are clear degeneracies in the observables represented by a given climate state. Note that we have not included any measurement errors, which would only exacerbate the problem.}\label{obfig:hrspecs}
\end{figure*}

The set of PlaSim experiments used in our model ensemble are summarized in \autoref{obtable:models}. Most of the models in the ensemble come from large multi-dimensional parameter sweeps involving rotation rate, fully-frozen initial conditions (to build a large ensemble of snowball models), and land fraction and distribution. All models used warm-start initial conditions, with the exception of the cold-start snowball models. To save time, we did not re-run models that crashed or stalled, and simply excluded them from our final ensemble. These models account for approximately 1\% of our initial model ensemble, and are not included in the counts shown in \autoref{obtable:models}. We did not hold instellation fixed across all parameter sweeps, as we are interested primarily in how a variety of climate states appear in observables. Whereas real exoplanet climates are sensitive to a wide range of factors not varied in our models, such as topography \citep{Blumsack1971} and ocean salinity \citep{Cullum2016}, we ensure a diverse range of climates in our model bank by choosing the instellations most likely to produce both warm and cold climates. We normalize to instellation when comparing our spectra, such that we are primarily sensitive to reflectance, not instellation. 

In the case of rotation rate, we varied the length of the solar day from 4 hours to 100 days, and for each rotation rate we also varied the CO$_2$ partial pressure pCO$_2$ from 1 $\mu$bar to 0.1 bar, to capture a large range of sea ice fractions. Our ensemble includes 964 models from the rotation rate model grid. 

When we varied land fraction, we included 8 different land fractions ranging from 0.1 to 0.8. To marginalize over the effects of supercontinents and different mean continental latitudes, for each land fraction we generated 5 different land distributions using the procedure described in the next section. For each land distribution, we then varied pCO$_2$ from 1 $\mu$bar to 0.1 bar. We included 395 models from the land fraction experiment. The instellations for both of these parameter grids were chosen to ensure that the models with low pCO$_2$ were either fully-frozen or nearly fully-frozen, and the models with high pCO$_2$ were fully-temperate. 

In our cold-start parameter sweep, we started from PlaSim's default warm initial conditions, reduced the instellation by 25 W/m$^2$, and then further reduced the instellation by 25 W/m$^2$ each model year until the planet had a mean surface temperature below 230 K for a minimum of 30 years. We then ramped the instellation back up to its initially-prescribed value by 25 W/m$^2$ per year. Once the instellation was back to its prescribed value, the model was allowed to run to equilibrium, for a minimum of 50 additional years. Our cold-start parameter sweep sampled CO$_2$ partial pressures ranging from 1 $\mu$bar to 10 mbar and instellations ranging from 800 to 1400 W/m$^2$. Our ensemble includes 299 models from the snowball parameter grid. In cases where pCO$_2$ was varied, we kept the N$_2$ partial pressure pN$_2$ constant, such that the total surface pressure increased with elevated pCO$_2$.

We included spectra for the models in the Earth-like pN$_2$ grid we described in \citet{Paradise2021}, specifically Figure 2. In these models, we varied pN$_2$ from 0.1 atm to 10 atm and instellation from 1100 to 1550 W/m$^2$, using Earth-like land and obliquity, to include the effects of varying amounts of Rayleigh scattering and the range of climates found at different background gas pressures. We did not include obliquity in our synthetic spectra, such that our spectra correspond to an equinox-like zenith angle. We included 179 spectra from this grid. In addition, we also ran several smaller experiments. We varied pN$_2$ from 1 atm to 10 atm for an aquaplanet configuration (zero land), including 10 models to account for the role of elevated Rayleigh scattering without the effect of land. We also included 20 models with varying pCO$_2$ for an aquaplanet at nearly Earth-like instellation, to include warmer climates with elevated CO$_2$ levels. Finally, we include 8 `desert planet' models, where we initialize the model with no surface oceans, no soil water, and zero humidity. For these models, we set a uniform surface albedo ranging from 0.1 to 0.8, and chose an instellation for each that gives a 220 K effective temperature:

In addition to these large parameter sweeps, we also included a few smaller experiments. We varied pN$_2$ for a planet completely covered in ocean, including 10 models as a comparison with the larger sweep, which used Earth continents. We also included 20 aquaplanet models with varying CO$_2$ partial pressure, all with warm-start initial conditions at 1350 W/m$^2$, in order to vary emissivity among the temperate models. Finally, we also included 8 cold `desert planet' models---we specified a uniform land surface, with no groundwater, and no atmospheric water, in order to test the extent to which a cold, dry planet could mimic either a snowball planet or a temperate planet. For these models, we set a uniform surface albedo ranging from 0.1 (dark soil) to 0.8 (white or reflective rocks or minerals, such as quartz or calcite; \citet{aster,ecostress}), and choose an instellation for each that gives a 220 K effective temperature:
\begin{linenomath*}
\begin{equation}
S = \frac{4\sigma T_\text{eff}^4}{1-a}
\end{equation}
\end{linenomath*}
where $S$ is the instellation of the planet, $T_\text{eff}$ is the effective temperature (220 K in this case), $a$ is the uniform surface albedo, and $\sigma$ is the Stefan-Boltzmann constant. 

\subsubsection{Randomized Continents}\label{obsec:land}
\begin{figure*}
\begin{center}
\includegraphics[width=5in]{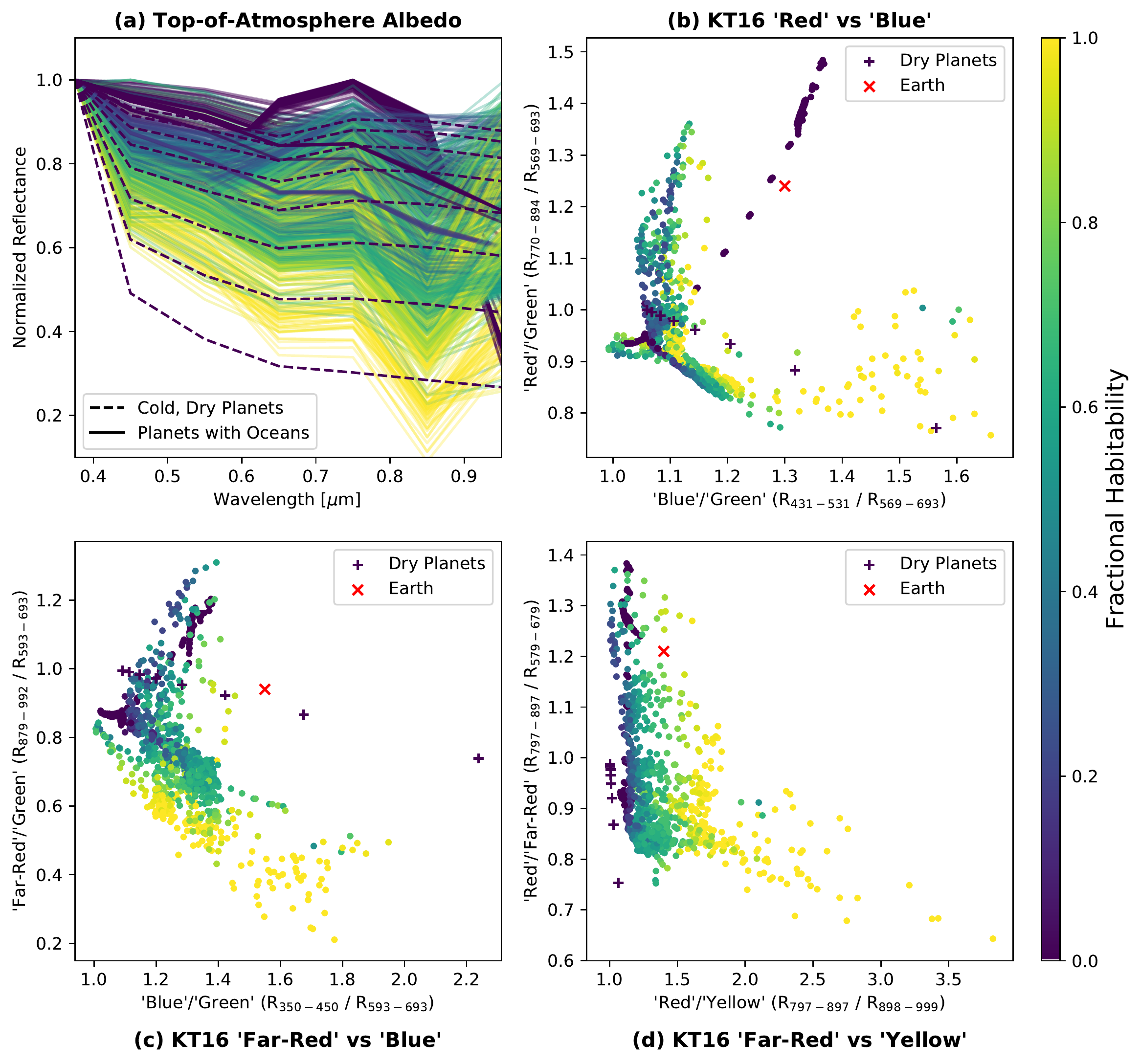}
\end{center}
\caption{The observables reported in \citet{Cowan2011} and \citet[][``KT16"]{Krissansen-Totton2016}, computed for our considerably larger ensemble of models. Panel (a) shows normalized reflectivities, computed using the same bins as \citet{Cowan2011}. Normalization breaks the degeneracy between planet radius and albedo. Panels (b), (c), and (d) show our models' positions in the three color-color spaces identified by \citet{Krissansen-Totton2016} as most-likely to separate Earth-like planets from false positives. Cold, dry planets are planets lacking water, with snowball-like temperatures, and which can have a range of surface reflectivities. Earth remains easily-separable in the red/blue color space because our models do not include the significant Chappuis O$_3$ absorption \citet{Krissansen-Totton2016} identified as the cause for the separation in that color space. In general, however, the overlap between bright and dark symbols suggests that fractional habitability cannot always be easily inferred with these techniques without additional constraints. Note that no measurement errors have been included.}\label{obfig:viscompar}
\end{figure*}

To explore the effects of land fraction and distribution, we implemented an algorithm to randomly distribute and build continents, as this is how land is distributed on Earth, as opposed to a fully-uniform distribution. This algorithm begins by randomly selecting points to be `continent seeds', and then using rejection sampling to assemble continents attached to those grid cells. We use periodic boundary conditions in longitude, and a variant of a symmetric boundary condition in latitude, where the row of latitudes representing the polar boundary is flipped, so that each cell can `see' the cells directly opposite the pole. Our continent-generation code is included with the \href{https://doi.org/10.5281/zenodo.2602241}{version of PlaSim available on AP's GitHub}, and the algorithm is as follows:
\begin{enumerate}
\item Draw $n$ random cells from a list of all grid cells, weighted by their areas. These are the `continent seeds', and are assigned a value of 1.
\item While the land fraction is less than the target value $f_\text{land}$, randomly draw a longitude from $\phi\in[0,2\pi]$ and a latitude from $\theta=\arcsin(1-2u)$, where $u\in[0,1]$:
\begin{itemize}
    \item If the corresponding cell has land already, reject it.
    \item If none of the neighboring 8 cells have land, reject it (only keep cells which will build on to existing land).
    \item If fewer than 5 of the 8 neighboring cells have land, reject the chosen cell 90\% of the time. This creates a preference for contiguous land areas, rather than many inland seas.
    \item If the chosen cell is not rejected for the above reasons, assign a value of 1 to indicate that it has land.
\end{itemize}
\end{enumerate}
This algorithm converges to the desired land fraction $f_\text{land}$ in T21 resolution relatively quickly, avoids continents with many inland seas (preferring contiguous land areas similar to Earth's continents), and avoids the projection-related shape distortions that come with simply uniformly-sampling in latitude and longitude. Examples of the land distributions produced are shown in \autoref{obfig:continents}. While the algorithm begins with $n$ continent seeds, in most cases fewer than $n$ discrete continents are produced, as land areas can merge as they grow, in some cases producing supercontinents.

\section{Results \& Discussion}

\begin{figure*}
\begin{center}
\includegraphics[width=6.5in]{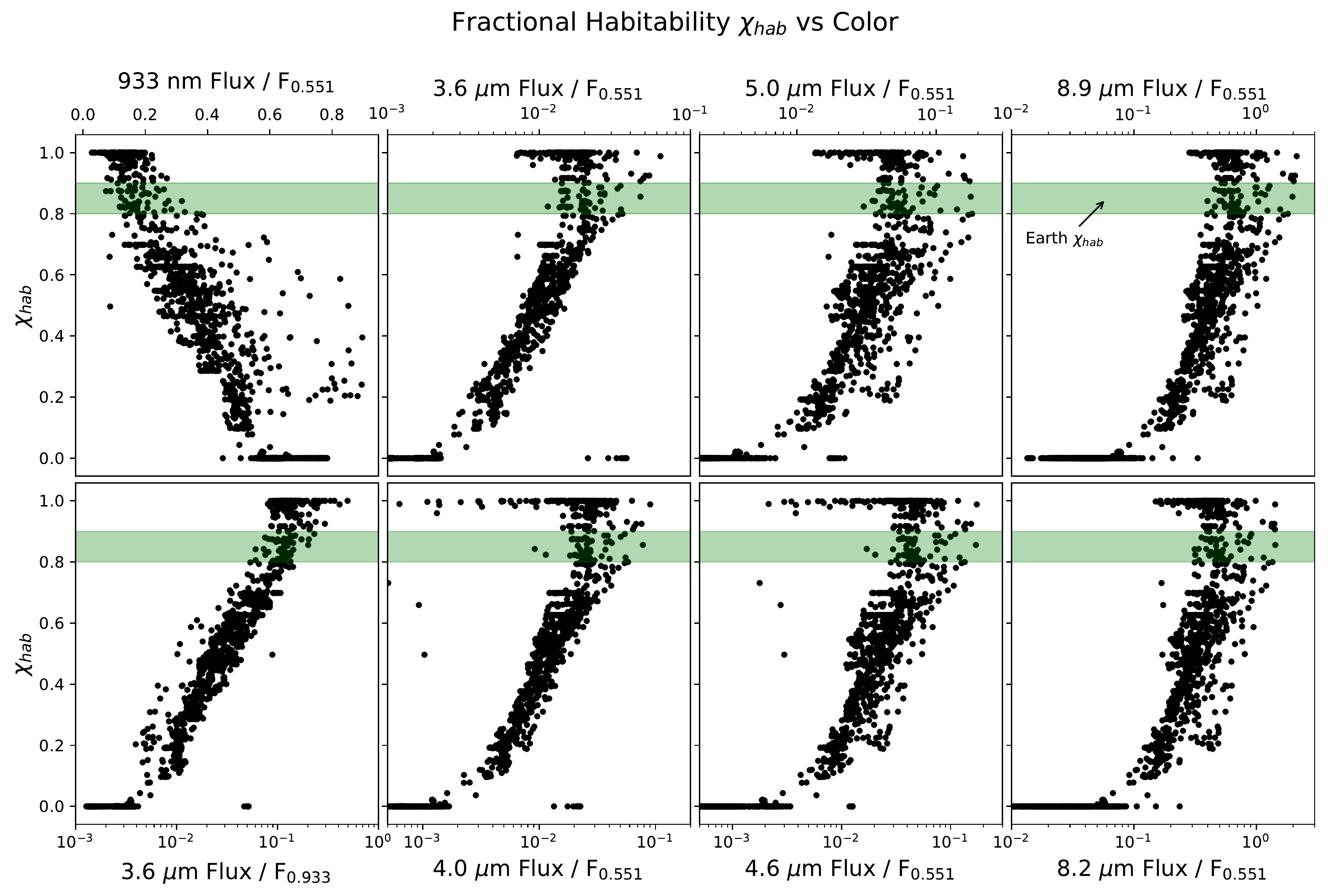}
\end{center}
\caption{Here we show annual average fractional habitability $\chi_\text{hab}$ as a function of 8 different colors identified as having either the strongest correlation (the top row) or covariance (the three rightmost panels of the bottom row) with fractional habitability. The seasonal $\chi_\text{hab}$ range for modern Earth \citep{Spiegel2008} is shown as a shaded green region. Note that we have not systematically varied obliquity in our model bank. The bottom-left panel uses the color constructed from the two wavelengths with the highest correlation coefficients. No measurement errors have been included. The cold desert planets can be seen as clear outliers, with colors similar to their temperate counterparts. While there are clear trends that will permit statistical inference from surveys of the habitable exoplanet population, even these optimal colors show significant degeneracies in the climates associated with a given observable.}\label{obfig:colors}
\end{figure*}

\begin{figure*}
\begin{center}
\includegraphics[width=6.5in]{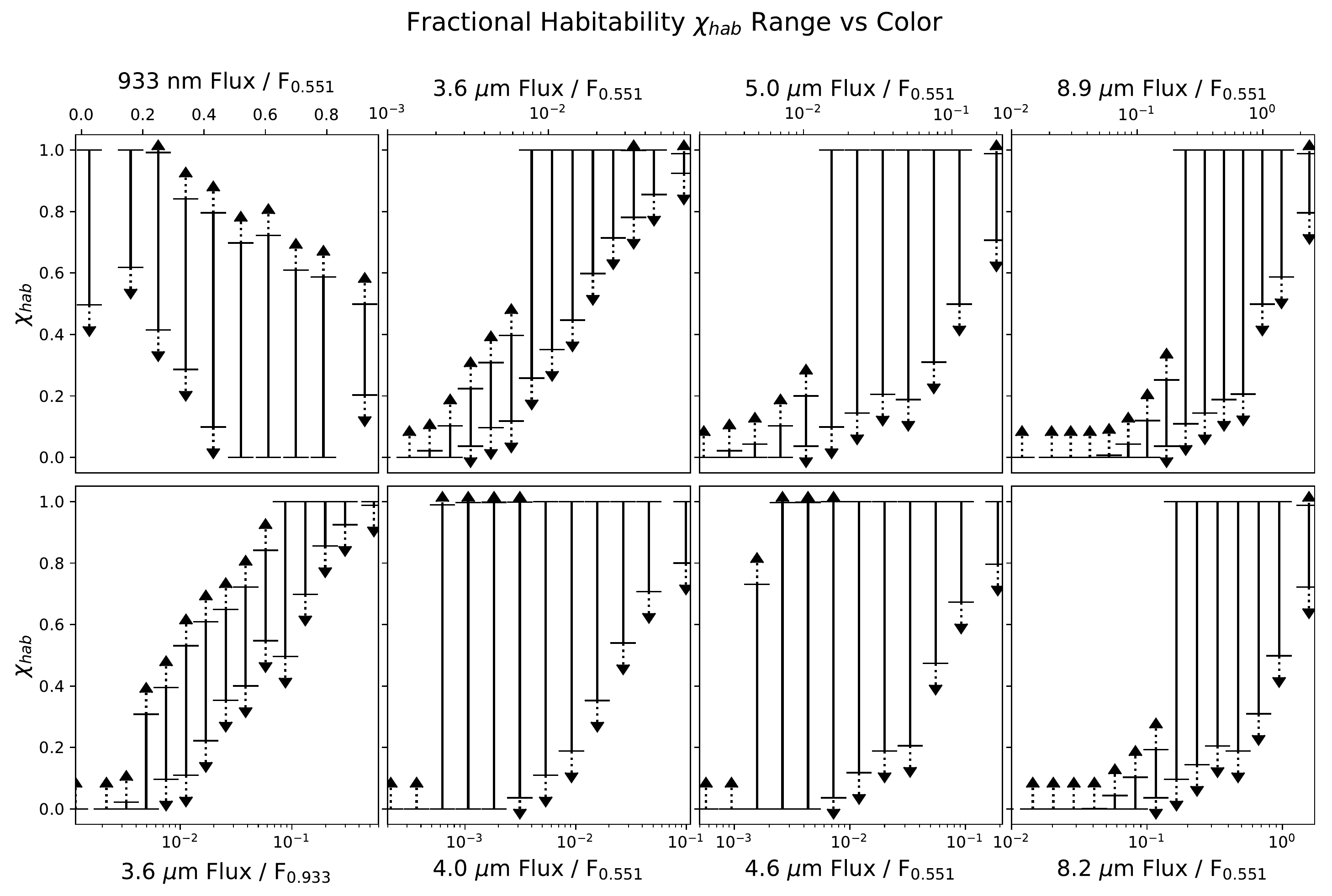}
\end{center}
\caption{The approximate range of minimum uncertainty on fractional habitability $\chi_\text{hab}$ observed in our sample for each of the 8 colors we identified. The arrows indicate that these ranges represent minimum bounds on the range of $\chi_\text{hab}$ that may produce a certain color brightness, and further studies with a broader diversity of climates may in fact increase these ranges, increasing the uncertainty of climate retrievals. Note that these are not statistical error-bars, but instead denote the maximum observed variance in $\chi_\text{hab}$ within our sample as a function of color intensity, and thus the minimum uncertainty that can at present be assumed for a hypothetical observation (with no prior constraints on the underlying population). In this plot, we omit individual markers for individual climates, and instead show only the approximate bounds of the data to emphasize the point that our models are not statistically representative of the true distribution of Earth-like climates, and apparently-tight correlations observed within our sample may turn out to be artifacts of our parameter choices and model assumptions.}\label{obfig:colorsrange}
\end{figure*}

\begin{figure*}
\begin{center}
\includegraphics[width=6in]{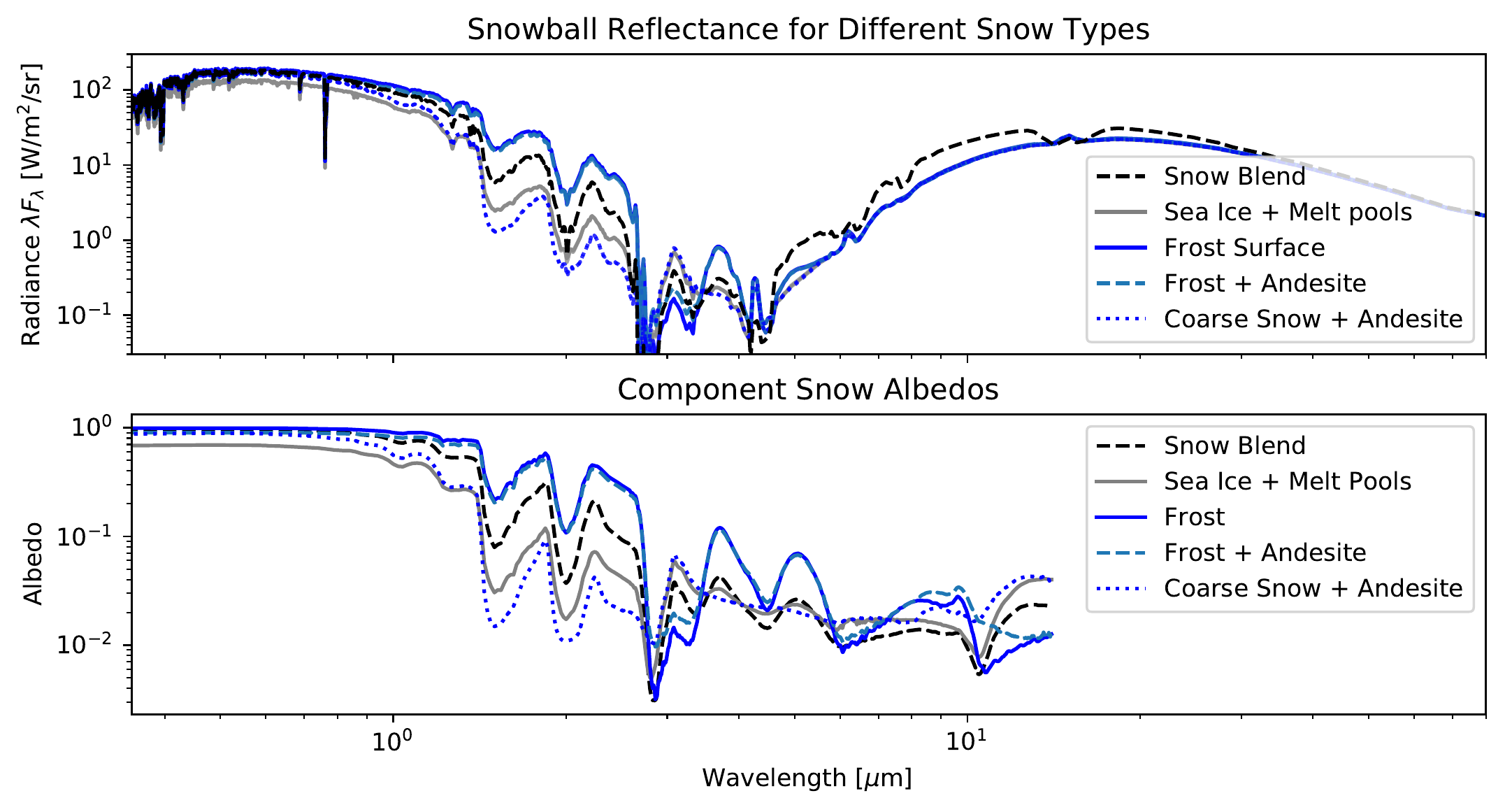}
\end{center}
\caption{The top panel shows radiances for several snowball models re-computed with different spectral snow models, derived from samples from the JPL ECOSTRESS spectral library \citep{aster,ecostress}; corresponding albedo spectra are shown in the bottom panel. No measurement error is included. The 'snow blend' is the model used in the rest of our analysis; the others reflect surface types which would be consistent with ``snow" or ``ice" surfaces in PlaSim, and whose existence and prevalence on exoplanet surfaces would be difficult to constrain. The variance in observables shown here due to different types of `icy' surfaces is not included in the rest of our spectral analysis, and illustrates the extent to which inference is sensitive to model assumptions.}\label{obfig:iceblends}
\end{figure*}

The synthetic spectra from our postprocessed PlaSim models are shown in \autoref{obfig:hrspecs}. We stress here that spectral data of this quality is not possible with existing or upcoming observational facilities. Rather than assess the techniques that will be accessible with such facilities, our goal here is to identify the fundamental challenges that will drive improvements to modeling and retrieval techniques in the coming decades.

We first re-examined the techniques described in \citet{Cowan2011} and \citet{Krissansen-Totton2016}, recomputing the observables they explored. The assumption underlying both these techniques and our motivating question is that cold planets, covered in reflective ice, ought to look different from warmer planets with darker oceans. Our results are shown in \autoref{obfig:viscompar}, and suggest that while the extremes, such as very warm temperate planets and very cold snowball planets, can be easily-distinguished with both techniques, it is challenging to differentiate partially-glaciated planets and comparatively warmer snowball planets. The possibility of cold planets that lack water and yet have reflective surfaces can further confound this inference. Earth remains separable in color-color space only because our spectra did not include the ozone Chappuis bands responsible for Earth's comparative blue color \citep{Krissansen-Totton2016}. 

Both of these techniques \citep{Cowan2011,Krissansen-Totton2016} were however limited to wavelengths less than 1 $\mu$m. The argument that cold planets should look different from warm planets can be reframed as an energy budget argument---a planet that reflects a lot of light and has little thermal emission is probably cold, while a planet that absorbs a lot of light and emits more energy in the infrared is probably warm. We therefore considered a broader spectral range extending out to 80 $\mu$m, sorted the spectra in order of increasing fractional habitability $\chi_\text{hab}$ (middle panel of \autoref{obfig:hrspecs}), and computed the correlation coefficient between $\chi_\text{hab}$ and the logarithm of the radiance relative to the radiance at 551 nm, as shown in the bottom panel of \autoref{obfig:hrspecs}. We also considered the wavelengths with the highest correlation coefficients using the radiance itself, rather than its logarithm, but found that the correlation coefficients of those wavelengths were lower than those found with the logarithm of the radiance, and manual inspection showed that these wavelengths were generally not useful for inference due to large scatter around any potential trends. We expect the wavelengths with the highest correlation coefficients to have a relatively monotonic relationship between normalized radiance and fractional habitability, indicating their suitability for observational climate retrievals.

We select wavelengths that locally-maximize the correlation between log normalized radiance and $\chi_\text{hab}$, as this may allow us to combine distinct wavelengths that probe different physical processes in the atmospheres of the models in our sample. We identify 933 nm, 3.6 $\mu$m, 5 $\mu$m, and 8.9 $\mu$m as wavelengths at which fractional habitability is most-strongly correlated with the brightness of the planet. With the exception of 933 nm, which is in a water absorption band, these wavelengths are near the edges of window bands, which allows them to both probe the surface and near-surface atmospheric layers, as well as changes in radiatively-active absorbers such as CO$_2$ and water. We also identify three wavelengths in the infrared at which the covariance of the logarithm of the radiance and fractional habitability are maximized: 4 $\mu$m, 4.6 $\mu$m, and 8.2 $\mu$m. If these wavelengths prove suitable, their higher covariances would indicate lower photometric sensitivity requirements. The high-covariance wavelengths are on the edges of window bands just like the high-R$^2$ wavelengths. The core of our analysis however focuses on correlation rather than covariance; we are interested primarily in how well a climate can be constrained with specific observables, rather than primarily reducing observational expense. We note that the fact that the bands we identify as optimal differ from those in \citet{Krissansen-Totton2016} suggests that the choice of optimal observing band may itself be model-dependent. Our approach to identifying the `best' wavelengths may also select wavelengths which are dim, and therefore more challenging to observe, but we reiterate that our aim is to explore the fundamental obstacles preventing confident climate retrievals, even in the limit of ideal observations in the most-informative wavelengths.

We examined the relationship between radiance at these wavelengths and fractional habitability, as shown in \autoref{obfig:colors}. The seasonal range of fractional habitabilities found on Earth, 0.8--0.9 \citep{Spiegel2008}, is shown in the figure as a shaded green region. We normalize each radiance to the radiance at 551 nm to remove the degeneracy between radius and albedo when calculating the disk averaged brightness. This normalization to a visible wavelength also allows us to assess the energy budget, by comparing emission to reflection. There are clear trends in each case, and as in \autoref{obfig:viscompar} the extremes are easily-identified--i.e., at a certain point the energy budget requirement dominates. However, there is a significant amount of scatter for climates between the hot and cold extremes. Note as well that obliquity and asymmetric land distribution, as on Earth, can result in seasonal variations in $\chi_\text{hab}$. Our models with Earth continents had Earth obliquity, but all the others had zero obliquity---obliquity was not a parameter we systematically varied in our model bank. The scatter shown here therefore means that while statistical analyses of the habitable exoplanet population are in principle feasible, inferring the individual climates of the most Earth-like planets requires that retrieval frameworks sample a very large parameter space.

To explore the possibility that inference may simply be under-determined with only one color, we also examine the ratio of the 3.6 $\mu$m and 933 nm radiances, as shown in the bottom-left panel of \autoref{obfig:colors}. These two wavelengths probe surface reflectivity and atmospheric water absorption respectively, and so together may provide tighter constraints. This does reduce the scatter somewhat, producing our strongest correlation with fractional habitability, and suggesting that future statistical surveys of Earth-like exoplanet climates may observe a `main sequence' of habitable climates, where habitability (or surface temperature) is correlated with color or a combination of colors, similar to how stellar properties are correlated with color and brightness. Even in this case, however, among temperate planets $\chi_\text{hab}$ varies by 0.2--0.4 or more for a given color brightness, and we are not able to reduce the scatter further. In particular, we note that it may not be possible to distinguish a planet with small habitable areas from a fully-frozen planet, nor a planet with 40\% ice coverage from a planet completely lacking sea ice. Outliers to the main trends seens in each panel contribute to this uncertainty. While it is possible that combining several colors into a higher-dimensional color space would further reduce the scatter to the extent that a confident inference could be made, using more than one color dramatically increases the SNR required at each wavelength to make a $5\sigma$ inference \citep{Krissansen-Totton2016}. Our results suggest that at least three colors are needed, making such an approach most likely unfeasible without additional observational constraints.

More generally, while the climate parameter space surveyed in \autoref{obfig:colors} is extensive, it is by no means exhaustive. A key challenge to any reliable climate inference in the future will thus be the need to confidently rule out the possibility of any `outlier climate' in this type of color scatter plot. In particular, we stress here that because the actual prevalence of different surface types, atmospheres, and climate states on Earth-like planets is unconstrained, a single `outlier climate' can greatly increase the uncertainty of a retrieval. To illustrate this, in \autoref{obfig:colorsrange} we have plotted only the ranges of fractional habitability corresponding to each color, omitting the cold desert climates (under the assumption that the lack of water vapor could be used to remove them from a sample with follow-up observations). Note that these are not statistical error-bars representing 1-sigma uncertainties, but instead represent the maximum observed variance within the sample for each $\chi_\text{hab}$ as a function of color intensity. Without prior constraints on the underlying distribution, this translates to a minimum uncertainty that can be assumed. Because our sample is not statistically representative of the true distribution of Earth-like exoplanet climates, any apparently-tight correlation in our sample may be an illusion caused by our parameter choices and model assumptions. We must therefore limit ourselves to the minimum uncertainty range implied by the data, and caution that further studies with a broader diversity of climates may widen this range. Further theoretical work focusing on outlier climates to predict their occurrence rate and search for ways to rule them out is needed for any attempt to narrow this range. Given that Earth-like priors typically characterize studies of habitable climates, it remains to be seen whether or not confidently ruling out outlier climates will be possible on the same timescales as upcoming exoplanet characterization efforts.

\section{Additional Uncertainty}

Much of the variance in our observables is due to variations in cloud coverage. Clouds, a common source of disagreement in models \citep{cam5clouds,Yang2019,Fauchez2020}, make the planet more reflective in the visible and may in the case of high-altitude clouds mask surface thermal emission. Rotation rate, land fraction, and surface pressure all affect the cloud fraction found in our models for a given fractional habitability, and we note that cloud cover is also sensitive to factors we did not vary, such as aerosols and airborn particulate matter \citep{Cloudreview,cam5clouds,cmip2019}. The variance in our models' spectra due to clouds is therefore likely an underestimate of the true sensitivity to a model's cloud parameterization, and its indirect sensitivity to assumptions such as land fraction and rotation rate.

Our observables are also directly sensitive to a range of model assumptions that are often treated as free parameters, including land fraction and configuration, atmospheric mass, and rotation rate. Cold, icy planets with large land areas do not necessarily have uniform snow and ice cover on land \citep{Paradise2019}, and may appear relatively dark in visible wavelengths despite globally cold temperatures. Differences in heat transport due to rotation rate \citep{Showman2014} or atmospheric mass \citep{Vladilo2013} can lead two climates with the same fractional habitability to have different mean surface temperatures and temperature distributions, resulting in changes to synthetic observables.

One of the biggest sources of uncertainty in the spectra in \autoref{obfig:colors} is the set of reflectance models we assume for various surface types. Our SBDART configuration includes only a few surface reflectance models---sand, water, and snow, with our darker land surfaces being simple weighted combinations of sand and water \citep{disort,Ricchiazzi1998,modtran}. We explored the importance of these assumptions by recomputing several snowball model spectra with different assumed snow reflectance models, computed as weighted combinations of reflectance models from the ECOSTRESS spectral library \citep{aster,ecostress}. We considered four reflectance models in addition to our default: sea ice plus melt pools, a frost-covered surface, a mix of frost and exposed rock, and a mix of coarse snow and exposed rock. Each would be considered uniform snow or ice coverage in PlaSim. As shown in \autoref{obfig:iceblends}, the choice of surface reflectance model for a given surface type can have a large impact on simulated near-infrared observables, adding an additional source of model-dependence for retrievals.  


We have not considered observations of our planets at multiple phase angles, nor have we considered spectropolarimetry, both of which may help to constrain surface conditions \citep{Cowan2013,Fujii2017,Trees2019,Wolf2019}. Ocean glint in particular may be detectable through either photometry or spectropolarimetry at certain phase angles \citep{McCullough2006,Stam2008,Williams2008,Robinson2010,Visser2015,Emde2017,Lustig-Yaeger2018}, and may help to identify climates with temperate surface conditions. It is however likely that there are additional sources of model-dependence not included in our experiment which will further increase the difficulty of infering planetary surface conditions, similar to the specific types of `icy' surfaces explored above. Advanced techniques such as spectropolarimetry and time-resolved or phase-resolved spectroscopic observations are still in early stages of investigation, and have yet to be applied to very large samples of 3D models such as ours. Our climate model outputs in addition to our spectra are publicly-available to facilitate further study in this direction. While planet-specific inferences may prove challenging or unfeasible with the current state of climate models, we do find several clear correlations between fractional habitability and various observables, lending support to the argument for a statistical approach to the search for habitable planets \citep{Bean2017,Checlair2019}, and motivating further research into statistical observables using large model ensembles.

We note once again that the range of climates included in our sample is not exhaustive, and is in fact quite limited, as it represents only Earth-like atmospheres that can be modeled by our modified version of PlaSim. In reality, an enormous diversity of atmospheres beyond that parameter space exists, including hazy worlds such as Titan and Archaean Earth, CO$_2$-dominated atmospheres such as Mars, H/He-dominated atmospheres, and runaway greenhouse and post-runaway greenhouse climates such as Venus. We have omitted these climates primarily because PlaSim cannot reliably model them, and we are able to capture a large, computationally-expensive parameter space nonetheless with even our small number of varied parameters. We also however omit such climates from our analysis because as atmospheres deviate more from an Earth-like atmosphere, it may be increasingly easy to identify them as non-Earth-like, even if they cannot be ruled uninhabitable or non-temperate. Those that cannot be easily-differentiated would only serve to reinforce our conclusions, by widening the range of climates and fractional habitabilities represented by a given observable. We have therefore kept our focus relatively narrow, to solely Earth-like climates, which allows us to set minimum bounds on the fundamental uncertainties exoplanet scientists are likely to encounter when observing potentially Earth-like planets.

Finally, we note that we have omitted uncertainties due to instrumentation and measurement errors. Real instruments have finite signal-to-noise, and may be limited in the magnitudes that can be observed at a given wavelength, either by their design or by limited observation time. We have omitted this (often large) contribution to the problem of inferring climate for two reasons. First, we wish to explore fundamental uncertainties that result from variance in exoplanet climates themselves. The uncertainties and variance shown here are entirely separate from instrumentation noise, and will be present even with perfect observations with infinite SNR. Put another way, instrumental uncertainties would appear in \autoref{obfig:colors} as horizontal error bars, whereas the uncertainty we are highlighting should be interpreted as \textit{vertical} error bars (as in \autoref{obfig:colorsrange}, but again note that these are minimum uncertainties being presented, not expressions of the underlying population's statistical distribution). Second, we believe that instrumentation-based barriers to inference are temporary. Astronomical instrumentation has improved rapidly, and it is likely that any assumptions we make about the ability of a given observing platform to make an observation will be soon outdated as new data reduction techniques emerge and improvements to instruments are devised. The problem that we have focused on in this study, however, is not temporary---or at least, the variance of the exoplanet climate distribution will not change. Resolving this problem will require better knowledge of exoplanet climate dynamics and new inference and observation techniques.

\section{Conclusion}

In summary, we have examined high-resolution spectra of reflected and emitted light spanning visible and infrared wavelengths for nearly two thousand 3D climate models, and found that even in a very limited slice of the possible Earth-like exoplanet parameter space, there is substantial variance in simulated spectra, indicating that future retrieval efforts will likely be highly-sensitive to basic model assumptions. Our model outputs and spectra, representing approximately 200,000 years of climate and over 100 GB of data, are available for further community analysis as noted in the Data Availability section below. At present, it is unfeasible to account for all relevant assumptions and parameterizations in a 3D retrieval framework. It is possible that combining full-phase reflection and emission spectra with other types of observations, such as transmission spectroscopy, temporal phase curves, eclipse mapping, spectropolarimetry, or detection of ocean glint may help to break the degeneracies we have observed, and this ought to be explored in future studies---but a multi-technique approach to climate inference significantly increases the observational cost of inferring individual climates. With full-phase spectra alone, even in the best-case scenario of color-color comparisons that use the wavelengths most strongly-correlated with fractional habitability, it will be extremely difficult to place tight constraints on the fraction of an individual planet's surface that is above freezing without significant improvements to the modeling hierarchy and a thorough exploration of the possible climate parameter space. The fundamental challenge facing the field is that this exploration relies on model priors that are Earth-centric and whose performance on climates other than Earth's has not yet been tested. Our results do suggest that broad correlations may exist, which may be possible to confirm observationally through large, stastical surveys of Earth-sized exoplanets in the habitable zone, and we encourage further study of possible statistical trends in our data using our publicly-available model bank and spectra.

\section*{Acknowledgements}

AP is supported by the Department of Astronomy \& Astrophysics at the University of Toronto, and by the Province of Ontario through the Ontario Graduate Scholarship Program. KM is supported by the Natural Sciences and Engineering Research Council of Canada. CL is supported by the Faculty of Arts \& Science Tri-Council Bridge Funding and the Department of Physics at the University of Toronto. Computing resources were provided by the Canadian Institute for Theoretical Astrophysics at the University of Toronto (CITA). The version of PlaSim used in this study is available on AP's GitHub \citep[v1.1.0]{gplasim}. We are grateful to the anonymous reviewer for their helpful comments, and to Jean Schneider and Tyler Robinson for their excellent comments and thoughts on ocean glint and spectropolarimetry as potential ways to break the degeneracies we have noted.

We would furthermore like to acknowledge that our work was performed on land traditionally inhabited by the Wendat, the Anishnaabeg, Haudenosaunee, M\'{e}tis, and the Mississaugas of the Credit First Nation. Furthermore, this work required supercomputing infrastructure, whose availability was only possible through the mining of precious metals from ecologically-sensitive areas around the world that are traditionally inhabited by Indigenous peoples. Finally, supercomputing is energy-intensive, and often therefore carbon-intensive. Power generated in Ontario and used by CITA comes from renewable and low-carbon sources \citep{CER}, but we acknowledge that this may not be true of follow-up work.

\section*{Data Availability}\label{sec:data}

The \href{https://github.com/alphaparrot/ExoPlaSim/releases/tag/v1.1.0}{version of PlaSim used in this study} is available on AP's GitHub \citep[version 1.1.0,][]{gplasim}, as well as through the Python Package Index via \texttt{pip install exoplasim}. Documentation is available online at \url{https://exoplasim.readthedocs.io/en/latest/}. Our model bank and the associated postprocessed spectra are available through an online \href{https://dataverse.scholarsportal.info/dataverse/kmenou}{Dataverse repository} \citep{Paradise2021b_data}. In addition to the spectra themselves, the repository contains the instantaneous model snapshots needed to recreate the spectra either through SBDART or an alternative radiative transfer code.



\bibliographystyle{mnras}








\bsp	
\label{lastpage}
\end{document}